\documentclass[prd,aps,superscriptaddress,floatfix,nofootinbib,eqsecnum,twocolumn]{revtex4-2}

\pdfoutput=1

\usepackage{amsmath}
\usepackage{amsfonts}
\usepackage{amsmath}
\usepackage{amssymb}
\usepackage{bm}
\usepackage{dcolumn}
\usepackage{graphicx}
\usepackage[utf8]{inputenc}
\usepackage{latexsym}
\usepackage{rotating}
\usepackage{textgreek}
\usepackage{mathrsfs}
\usepackage{hyperref}
\usepackage{subfigure}
\usepackage{color}
\usepackage{subcaption} 
\usepackage{graphicx}   
\usepackage{changes}
\usepackage{verbatim}
\usepackage{comment}
\usepackage{empheq}
\usepackage{csquotes}
\usepackage{physics}
\usepackage{xcolor}
\usepackage{float}
\usepackage{soul}
\usepackage{orcidlink}
\usepackage{amsmath,latexsym}
\usepackage{booktabs}  % For better horizontal rules

\begin{document}
\title{Photon angular momentum near Planck scale
}

\author{Kenil Solanki\orcidlink{0009-0007-0654-1616}}\email{d24ph006@phy.svnit.ac.in}\affiliation{Department of Physics \\ Sardar Vallabhbhai National Institute of Technology, \\ Surat, Gujarat,395007, India.}

\author{Gaurav Bhandari\orcidlink{0009-0001-1797-2821}}\email{bhandarigaurav1408@gmail.com; gauravbhandari23@lpu.in.}\affiliation{Department of Physics,\\ Lovely Professional University, Phagwara, Punjab, 144411, India}

\author{S. D. Pathak\orcidlink{0000-0001-9689-7577}}\email{sdpathak@lko.amity.edu; prince.pathak19@gmail.com}\affiliation{Amity School of Applied Sciences, \\ Amity University Uttar Pradesh, Lucknow Campus, Lucknow, 226028, India.}

\author{Vikash Kumar Ojha\orcidlink{0000-0002-0641-4015}}\email{vko@phy.svnit.ac.in}\affiliation{Department of Physics \\ Sardar Vallabhbhai National Institute of Technology, \\ Surat, Gujarat,395007, India.}

\begin{abstract}
We study the angular momentum structure of the gauge field in Lorentz covariant relativistic generalized uncertainty principle (RGUP) framework incorporating Planck scale minimal length effects. Using Noether's theorem for higher derivative RGUP-modified gauge field Lagrangian, we obtain the canonical and symmetric (Belinfante) energy-momentum tensors and the corresponding gauge spin and orbital angular momentum currents. We show that the canonical and Belinfante-Rosenfeld angular-momentum tensors continue to satisfy the standard conservation law in the presence of Planck-scale corrections. %These results support the stability of fundamental conservation laws under high-energy modifications.
The RGUP corrections introduce higher-order contributions to the angular momentum density and momentum flow, yielding a modified Poynting vector, with the Maxwell limit recovered for vanishing RGUP parameter.
\end{abstract}

%%%%%%%%%%%%%%%%%%%%%%%%%%%%%%%%%%
\maketitle

%%%%%%%%%%%%%%%%%%%%%%%%%%%%%%%%%%

%%%%%%%%%%%%%%%%%%%%%%%%%%%%%%%%%%

\section{Introduction }
A recurring theme across candidate theories of quantum gravity is the emergence of a finite, minimal measurable length scale, typically associated with the Planck length. This feature arises naturally in a wide range of approaches, including string theory (ST) \cite{corcoran1974string,gross1988string,tong2009string,kiritsis2019string}, loop quantum gravity (LQG) \cite{alfaro2002loop,chiou2015loop,rovelli2008loop,Ashtekar:2007tv,han2007fundamental}, noncommutative geometry (NCG) \cite{seiberg1999string,szabo2003quantum}, and doubly special relativity (DSR) \cite{kowalski2002doubly,Santamaria-Sanz:2025jfo}. The presence of such a minimal length effectively acts as a soft ultraviolet regulator, leading to modifications of the canonical phase-space structure underlying quantum mechanics and quantum field theory, and providing a systematic effective description of quantum-gravitational corrections at energy scales well below the Planck scale.

A convenient and widely used phenomenological framework to incorporate a minimal length is obtained by modifying the standard Heisenberg uncertainty principle (HUP). In its conventional formulation, the HUP permits arbitrarily precise position measurements at the cost of unbounded momentum uncertainty. Quantum-gravitational considerations, however, suggest that quantum fluctuations of spacetime geometry  obstruct the simultaneous measurement of canonically conjugate variables, such as position and momentum, with arbitrary precision \cite{mead1964possible,adler1999gravity}. These effects can be captured by introducing a quadratic momentum-dependent correction to the uncertainty relation, dubbed the Generalized Uncertainty Principle (GUP)\cite{bang2006quantum,bruneton2017quantum,maggiore1993generalized}. The resulting deformation introduces additional, irreducible uncertainties in measurement processes, thereby enforcing a finite minimal length scale and deforming the underlying phase-space structure of quantum mechanics.

The extension of the GUP framework to relativistic settings presents nontrivial challenges, particularly concerning Lorentz invariance and the composition of momenta. Consistent relativistic generalizations have been developed in \cite{Todorinov:2018arx} which ensure the existence of a Lorentz-invariant minimal length while avoiding frame-dependent artifacts and resolving issues associated with naive relativistic extensions. In this context, a particularly well-motivated and model-independent formulation is provided by the Relativistic Generalized Uncertainty Principle (RGUP)\cite{quesne2006lorentz,todorinov2020relativistic}, wherein the canonical commutation relations are deformed in a manner that preserves Lorentz covariance and yields frame-independent phenomenological predictions.

%A particularly well-motivated and model-independent way to encode these effects is through a Relativistic Generalized Uncertainty Principle (RGUP), which deforms the canonical commutation relations while preserving Lorentz covariance. RGUP-type frameworks arise in diverse approaches to quantum gravity, including string theory\cite{scardigli1999generalized,scardigli2019deformation,corcoran1974string,gross1988string,tong2009string,kiritsis2019string}, noncommutative geometry\cite{seiberg1999string,szabo2003quantum,manolakos2019noncommutative}, and doubly-special relativity\cite{amelino2002doubly,kowalski2002doubly,judes2002conservation}, and provide a systematic effective description of minimal-length physics at energies well below the Planck scale.

While the implications of generalized uncertainty principles have been studied
extensively in non-relativistic quantum systems \cite{Das2008,Bushev2019,Scardigli2015}, their consistent implementation in relativistic quantum field theory remains an active area of investigation.
In particular, RGUP modifies the definition of momentum operators, dispersion
relations, and phase-space measures, and therefore has direct consequences for
the realization of spacetime symmetries.  Since conserved quantities such as
energy-momentum and angular momentum are generated by these symmetries through
Noether’s theorem states that any deformation of the underlying kinematics necessarily
induces corresponding modifications in the associated currents and generators.
Understanding how RGUP affects these conserved structures is therefore essential
for assessing the physical consistency and phenomenological implications of
minimal-length scenarios \cite{bosso2020quantum,bosso2021quantum,bhandari2025rgup,Bhandari:2024akz,Bhandari:2025sgl}.
Angular momentum is a particularly sensitive conserved quantity, as it depends explicitly on both position and momentum operators and thus directly reflects the interplay between spatial localization and field dynamics.  In gauge field theories, angular momentum further decomposes into orbital and spin components, whose separation is already subtle even in standard quantum electrodynamics (QED) \cite{liu2015angular,bialynicki2011canonical}.  Issues of gauge invariance, Lorentz covariance, and the distinction between canonical and Belinfante constructions make the formulation of angular momentum for gauge fields a nontrivial problem \cite{damski2021impact,fan2021gauge,liu2015angular}.  These conceptual subtleties are
expected to be amplified in the presence of RGUP deformations, where the canonical phase space structure itself is modified.

Angular momentum serves as a powerful diagnostic of fundamental dynamics, as exemplified by its central role in modern high-energy physics and, in particular, by the long-standing proton spin puzzle in quantum chromodynamics (QCD) \cite{efremov1991quark,bagchi1993remarks}. Polarized deep-inelastic scattering experiments have demonstrated that the intrinsic spins of quarks account for only a small fraction of the proton’s total spin, with the remainder distributed among gluon spin and
orbital angular momentum.  This unexpected result revealed that angular momentum in relativistic quantum field theory is a highly nontrivial, scale-dependent quantity, sensitive to both dynamical correlations and the underlying structure of the theory.

Recent measurements at the Relativistic Heavy Ion Collider\cite{muller2006results} have begun to constrain the gluon spin contribution, while the proposed Electron-Ion Collider
(EIC)\cite{antonioli2024european} is expected to provide unprecedented access to the orbital angular momentum of quarks and gluons at very short distance scales.  At such high
energies and small spatial resolutions, it is natural to ask whether
quantum-gravity-inspired modifications of kinematics could leave observable
imprints on angular-momentum observables.  In this sense, angular momentum
provides a unique window into possible deviations from the standard quantum field
theory, complementing more traditional probes based on energy spectra or total
cross sections.

The angular momentum of the Gauge field,
with its well-defined separation into spin (polarization) and orbital
components, plays a central role in radiation processes, structured light, and
polarization transport.  Moreover, precision measurements of atomic transition
rates, polarization-dependent effects, and high-angular-momentum gauge states
have reached sensitivities capable of constraining extremely small deviations
from standard electrodynamics \cite{kostelecky2009electrodynamics,kostelecky2002signals}.

In this work, we investigate the structure of angular momentum for the
Gauge field within the framework of the Relativistic Generalized
Uncertainty Principle.  We develop a fully relativistic and gauge-invariant
formulation of RGUP-modified gauge angular momentum, deriving both the canonical
and Belinfante energy-momentum tensors and the associated angular-momentum
currents.  Special attention is paid to the conservation laws, the role of
minimal-length corrections in the orbital and spin sectors, and the consistency
of the resulting generators with spacetime symmetries.  Our analysis clarifies how Planck-scale deformations propagate from modified commutation relations to observable angular-momentum densities, thereby establishing a concrete link between microscopic minimal-length physics and macroscopic polarization and angular-momentum transport. Beyond its formal significance, this work provides a theoretical foundation for
exploring possible experimental and observational signatures of RGUP effects in electromagnetic systems.  By situating gauge angular momentum at the interface between quantum field theory, gauge symmetry, and quantum-gravity-inspired kinematics, our results contribute to the broader effort to identify robust and testable consequences of minimal-length physics.\\

This article is organized 
as follows. In section  \ref{sec2}, we have discussed the background of RGUP and RGUP modified gauge field. In  section \ref{sec3} we derives the RGUP-modified canonical energy-momentum tensor (EMT) and constructs the corresponding canonical angular-momentum tensor, identifying the RGUP corrections to the orbital and spin contributions. Section\ref{sec4} develops the RGUP-modified {Belinfante} EMT and extracts the improved angular-momentum current, yielding a gauge invariant decomposition of gauge  and OAM in a minimal-length setting. Section\ref{sec5} computes the RGUP-corrected momentum density and Poynting vector, identifies the higher-derivative contributions responsible for modified energy and momentum transport, and interprets their physical significance. Section\ref{sec6}, we have discussed the conclusion and the possible future extension of this work.

\section{Background of RGUP}\label{sec2}
The existence of a fundamental minimal length is a generic prediction of several approaches to quantum gravity, and black-hole Gedanken experiments \cite{mead1964possible,maggiore1993generalized}. A phenomenologically convenient way to encode this feature is through a deformation of the canonical phase-space algebra, commonly known as the Generalized Uncertainty Principle (GUP) \cite{bosso2020quantum,kempf1995hilbert,Kempf:1994su,scardigli1999generalized,scardigli2003generalized}. However, most formulations of GUP are intrinsically non-relativistic \cite{kempf1995hilbert,hossenfelder2013minimal,Scardigli2015,scardigli2019deformation}, leading to conceptual and technical inconsistencies when applied to relativistic quantum systems and quantum field theories.

To overcome this limitation, a Lorentz-covariant extension known as Relativistic Generalized Uncertainty Principle has been proposed \cite{quesne2006lorentz,todorinov2020relativistic}. RGUP incorporates a fundamental length scale while preserving (or controllably deforming) Lorentz symmetry, thereby providing a consistent framework for relativistic and field-theoretic applications. In this approach, the standard phase-space structure is modified through deformed commutation relations between spacetime coordinates and four-momenta, leading to corrections in dispersion relations, relativistic wave equations, and conserved currents. 

Specifically, in the RGUP framework, the position and momentum operators satisfy the covariant deformed algebra \cite{bosso2020quantum},

\begin{equation}
[x^\mu, p^\nu]
=
i \hbar \Bigl[(1 - \gamma p^\rho p_\rho)\eta^{\mu\nu} + 2\gamma p^\mu p^\nu \Bigr],
\label{eq:RGUP-algebra}
\end{equation}
where $\gamma = \gamma_0/(M_{\mathrm{Pl}}c)^2$ is a Planck-suppressed parameter. This deformation implies the existence of a Lorentz-invariant minimal length and renders the physical variables $(x^\mu,p^\mu)$ noncanonical.

To preserve the standard structure of relativistic quantum theory, it is convenient to introduce auxiliary canonical variables $(x_0^\mu,p_0^\mu)$ satisfying
\begin{equation}
[x_0^\mu,p_0^\nu]=i\hbar \eta^{\mu\nu},
\qquad
p_0^\mu=-i\partial^\mu,
\end{equation}
and to express the physical momentum as a nonlinear function of $p_0^\mu$. In particular, one may choose
\begin{equation}
p^\mu = p_0^\mu (1+\gamma p_0^\rho p_{0\rho}),
\end{equation}
which reproduces the algebra \eqref{eq:RGUP-algebra} to quadratic order in momenta. The corresponding mass-shell condition becomes
\begin{equation}
p_0^\rho p_{0\rho}\left(1+2\gamma p_0^\sigma p_{0\sigma}+\gamma^2 p_0^\sigma p_{0\sigma}p_0^\tau p_{0\tau}\right) = -m^2,
\label{eq:RGUP-dispersion}
\end{equation}
which represents the RGUP-modified Klein-Gordon equation in momentum space.

In position space, this modification translates into higher-derivative corrections to the field equations. For the Gauge field, we assume that Lorentz invariance and $U(1)$ gauge invariance are preserved, and that the RGUP deformation modifies the wave operator in the same way as in the scalar case. The RGUP-corrected Maxwell equations are therefore taken to be
\begin{equation}
\partial_\mu F^{\mu\nu}
=
\partial_\mu \partial^\mu A^\nu
+
2\gamma\,\partial_\mu \partial^\mu \partial_\rho \partial^\rho A^\nu
=0.
\label{eq:RGUP-Maxwell}
\end{equation}

The standard gauge-invariant field strength tensor is defined as usual by
\begin{equation}
F_0^{\mu\nu} = \partial^\mu A^\nu - \partial^\nu A^\mu.
\end{equation}
To incorporate the RGUP corrections in a gauge-covariant way, we define the modified field strength
\begin{equation}
F^{\mu\nu}
=
F_0^{\mu\nu}
+
2\gamma\,\partial_\rho \partial^\rho F_0^{\mu\nu}.
\label{eq:RGUP-F}
\end{equation}
This definition ensures that $F^{\mu\nu}$ remains gauge invariant, since it is constructed entirely from $F_0^{\mu\nu}$ and Lorentz-covariant derivatives.

With this definition, the modified Maxwell equations can be written compactly as
\begin{equation}
\partial_\mu F^{\mu\nu}
=
\partial_\mu F_0^{\mu\nu}
+
2\gamma\,\partial_\rho \partial^\rho \partial_\mu F_0^{\mu\nu}
=0.
\end{equation}

The corresponding gauge-field Lagrangian density is then obtained by the natural generalization of the Maxwell form,
\begin{equation}
\mathcal{L}_A
=
-\frac{1}{4}F_{\mu\nu}F^{\mu\nu}.
\end{equation}
Expanding to first order in $\gamma$, one finds \cite{bosso2020quantum},
\begin{equation}
\mathcal{L}_A
=
-\frac{1}{4}F_{0\mu\nu}F_0^{\mu\nu}
-
\frac{\gamma}{2}
F_0^{\mu\nu}\partial_\rho\partial^\rho F_{0\mu\nu},
\label{eq:RGUP-Lagrangian}
\end{equation}
up to total derivatives. This Lagrangian is manifestly Lorentz invariant and gauge invariant, and reduces to the standard Maxwell theory in the limit $\gamma\to 0$.

Equations \eqref{eq:RGUP-Maxwell}-\eqref{eq:RGUP-Lagrangian} provide a consistent effective field-theoretic description of gauge dynamics in the presence of a Lorentz-invariant minimal length. They form the basis for analyzing how RGUP corrections modify conserved currents, energy-momentum tensors, and angular momentum densities of the Gauge field.
\subsection*{Conventions and Field Definitions}

Throughout this work we employ the mostly-minus Minkowski metric
$\eta_{\mu\nu}=\mathrm{diag}(+,-,-,-)$.
The Gauge field-strength tensor is defined by
$F^{\mu\nu}=\partial^{\mu}A^{\nu}-\partial^{\nu}A^{\mu}$.
With these conventions its components are related to the electric and magnetic
fields according to,
\begin{equation}
 F^{i0}=E^{i},\qquad F^{0i}=-E^{i},\qquad 
F^{ij}=-\epsilon^{ijk}B^{k}
\end{equation}
where $\epsilon^{ijk}$ is the three-dimensional Levi-Civita symbol.
These relations fix our sign conventions and will be used throughout the paper.

\section{The Canonical Energy-Momentum Tensor}\label{sec3}
All physical theories must respect the conservation of fundamental quantities such as energy, momentum, and angular momentum. Noether's theorems establish a profound connection between these conservation laws and underlying symmetries \cite{halder2018noether}.Noether’s first theorem applies to continuous global symmetries of the action and guarantees the existence of a corresponding conserved current, providing the theoretical basis for conservation of energy-momentum and angular momentum in field theory. By contrast, Noether’s second theorem addresses continuous local (gauge) symmetries and yields differential identities among the Euler-Lagrange equations, expressing the gauge redundancy of the theory rather than producing independent conserved currents. Consequently, while the first theorem encodes physical conservation laws, the second constrains the dynamical and constraint structure of gauge-invariant field theories.
Various quantum field theory literature employ Noether's second theorem rather than the first to derive the energy-momentum tensor (EMT). In these derivations, spacetime-dependent translations \(x \mapsto x + \xi(x)\) are considered\cite{freese2022noether}, and surface terms are discarded. Despite this, the resulting EMT for scalar fields coincides with the canonical form obtained via Noether's first theorem\cite{greiner2013field}. It is then commonly assumed that this result can be extended to vector and spinor fields in a similar manner.
To construct the canonical energy-momentum tensor, we follow the general structure of Noether’s theorem for continuous spacetime transformations\cite{greiner2013field}.  
%\subsection*{RGUP Lagrangian and Noether Framework}

To derive the canonical energy-momentum tensor and the associated angular momentum in the RGUP-modified gauge field, we begin with the modified structure of the Lagrangian in Eq.(\ref{eq:RGUP-Lagrangian}) as,
\begin{equation}\label{Lagrangian}
\mathcal{L}(A_\mu,\partial_\nu A_\mu,\Box\partial_\nu A_\mu)
=
-\frac14 F_{\mu\nu}F^{\mu\nu}
-\frac{\gamma}{2}F^{\mu\nu}\Box F_{\mu\nu},
\end{equation}
and the gauge field is described by the four-potential \(A_{\mu}(x)\), with Gauge field strength tensor, $
F_{\mu\nu}=\partial_\mu A_\nu-\partial_\nu A_\mu$.
The contribution of RGUP correction is explicitly in the Lagrangian with the contribution of d'Almbertian operator $\Box=\partial_\rho\partial^\rho$. The parameter $\gamma$ denotes the RGUP parameter and becomes relevant in regimes where quantum-gravitational effects cannot be neglected. 

The appearance of the term \(F^{\mu\nu}\, \Box F_{\mu\nu}\) renders the dynamics higher derivative in the sense of Ostrogradsky \cite{pons1989ostrogradski,de1998ostrogradski}. Consequently, the canonical Noether construction must be generalized to incorporate the dependence of the Lagrangian on second derivatives of the fields. This requires introducing generalized momenta corresponding to both \(\partial_\nu A_\mu\) and \(\partial_\rho\partial_\nu A_\mu\).

%To derive the canonical energy-momentum tensor and the associated angular momentum in the RGUP-modified gauge field, we begin with the modified structure of the Lagrangian as\begin{equation}\label{Lagrangian}\mathcal{L}(A_\mu,\partial_\nu A_\mu,\Box\partial_\nu A_\mu)=-\frac14 F_{\mu\nu}F^{\mu\nu}-\frac{\gamma}{2}F^{\mu\nu}\Box F_{\mu\nu},\end{equation}and the gauge field is described by the four-potential \(A_{\mu}(x)\), with Gauge field strength\[F_{\mu\nu}=\partial_\mu A_\nu-\partial_\nu A_\mu.\]The contribution of RGUP correction is explicitly in the Lagrangian with the contribution of d'Almbertian operator \(\Box=\partial_\rho\partial^\rho\). where $\gamma$ is the RGUP parameter wich is relevant when the quantum gravitational effects are relevant. The RGUP-modified Lagrangian introduces a higher-derivative contribution through the d'Alembertian operator term \(\Box=\partial_\rho\partial^\rho\) on teh elctromagentic field tensor.  The resulting Lagrangian density is\begin{equation}\mathcal{L}=\mathcal{L}(A_\mu,\partial_\nu A_\mu,\Box\partial_\nu A_\mu)=-\frac14 F_{\mu\nu}F^{\mu\nu}-\frac{\gamma}{2}F^{\mu\nu}\Box F_{\mu\nu},\end{equation}which depends explicitly on the field, its first derivatives, and on second derivatives arising from \(\Box F_{\mu\nu}\).  The presence of higher derivatives requires a generalized form of the Noether procedure.

The dynamics follow from the action
\begin{equation}
\mathcal{W}=\int d^{4}x\,\mathcal{L}(x),
\end{equation}
and we study its variation under an infinitesimal transformation of the coordinates,
\begin{equation}
x'_{\mu}=x_{\mu}+\delta x_{\mu},
\end{equation}
together with a variation of the gauge field,
\begin{equation}
A'_{\mu}(x')=A_{\mu}(x)+\delta A_{\mu}(x),
\end{equation}
which gives the corresponding change in the Lagrangian density as
\begin{equation}
\mathcal{L}'(x')=\mathcal{L}(x)+\delta\mathcal{L}(x).
\end{equation}
At this point, it is crucial to distinguish two distinct variations, namely {Field variation or total variation} \(\delta A_\mu(x)\), which represent the intrinsic change of the field at the same coordinate label \cite{freese2022noether}, and {Coordinate variation} \(\delta x_\mu\), meaning shifting the argument of the field.

The functional variation for a covariant vector field is defined as,
\begin{equation}
\widetilde{\delta}A_{\mu}(x)
\equiv
A'_{\mu}(x)-A_{\mu}(x),
\end{equation}
and expanding \(A'_\mu(x)\) about \(x'\) and keeping only the first-order terms gives
\begin{equation}
\widetilde{\delta}A_{\mu}
=
\Delta A_{\mu}
-
\delta x^{\lambda}\partial_{\lambda}A_{\mu}.
\label{eq:tilde-delta}
\end{equation}
This form is particularly convenient because \(\widetilde{\delta}\) commutes with derivatives such that\cite{greiner2013field},
\begin{equation}\label{functioanlvar}
\widetilde{\delta}(\partial_\nu A_\mu)=\partial_\nu(\widetilde{\delta}A_\mu),\qquad
\widetilde{\delta}(\Box\partial_\nu A_\mu)=\Box\partial_\nu(\widetilde{\delta}A_\mu).
\end{equation}

Since our RGUP-modified Lagrangian contains higher derivatives, we define the generalized functional derivatives as 
%To organize the variation of the higher-derivative Lagrangian, we introduce the generalized functional derivatives,
\begin{equation}
S^\mu=\frac{\partial\mathcal{L}}{\partial A_\mu},\qquad
Q^{\nu\mu}=\frac{\partial\mathcal{L}}{\partial(\partial_\nu A_\mu)},\qquad
R^{\nu\mu}=\frac{\partial\mathcal{L}}{\partial(\Box\partial_\nu A_\mu)}\label{DERIVATIVE HIGHER ORDER}.
\end{equation}
Using the chain rule together with Eq.~(\ref{functioanlvar}) the variation of the Lagrangian takes the form

\begin{equation}
\widetilde{\delta}\mathcal{L}
=
S^\mu\widetilde{\delta}A_\mu
+
Q^{\nu\mu}\,\partial_\nu(\widetilde{\delta}A_\mu)
+
R^{\nu\mu}\,\Box\partial_\nu(\widetilde{\delta}A_\mu).
\end{equation}

We rewrite this expression by integrating by parts locally and moving all derivatives off \(\widetilde{\delta}A_\mu\).  
The term containing \(Q^{\nu\mu}\) and $R^{\nu \mu}$ gives
\begin{align}\label{Qmunu}
Q^{\nu\mu}\,\partial_\nu(\widetilde{\delta}A_\mu)
&=
\partial_\nu\!\left(Q^{\nu\mu}\,\widetilde{\delta}A_\mu\right)
-
(\partial_\nu Q^{\nu\mu})\,\widetilde{\delta}A_\mu,
\\[8pt]
R^{\nu\mu}\,\Box\partial_\nu(\widetilde{\delta}A_\mu)
&=
\partial_\nu\!\left(
R^{\nu\mu}\,\Box\widetilde{\delta}A_\mu
\right)
-
\partial_\nu\!\left(
\partial_\rho R^{\nu\mu}\,\partial^\rho\widetilde{\delta}A_\mu
\right)
\nonumber \\[4pt]
&\quad
+
\partial_\nu\!\left(
\Box R^{\nu\mu}\,\widetilde{\delta}A_\mu
\right)
-
(\partial_\nu\Box R^{\nu\mu})\,\widetilde{\delta}A_\mu
\label{rmunu}
\end{align}
After collecting all contributions from Eqs.~(\ref{Qmunu}) and~(\ref{rmunu}), 
we obtain the corresponding higher-derivative Noether identity

\begin{align}
\widetilde{\delta}\mathcal{L}
&=\left(S^\mu-\partial_\nu Q^{\nu\mu}-\partial_\nu\Box R^{\nu\mu}\right)\widetilde{\delta}A_\mu
\nonumber \\ &+\partial_\nu\Big[
Q^{\nu\mu}\widetilde{\delta}A_{\mu}
+R^{\nu\mu}\Box\widetilde{\delta}A_{\mu}
-\partial_{\rho}R^{\nu\mu}\partial^{\rho}\widetilde{\delta}A_{\mu}
+\Box R^{\nu\mu}\widetilde{\delta}A_{\mu}
\Big],
\label{eq:deltaL-final}
\end{align}
the first term in the above Eq.(\ref{eq:deltaL-final}) yields the higher-order Euler-Lagrange equation\cite{de1998ostrogradski,todorinov2020relativistic}, whereas the total divergence term identifies the associated Noether current.

For the RGUP-modified Lagrangian, one finds
\begin{equation}\label{emt}
S^\mu=0,\qquad
Q^{\nu\mu}=-F^{\nu\mu}-\gamma\,\Box F^{\nu\mu},\qquad
R^{\nu\mu}=-\gamma F^{\nu\mu}
\end{equation}
and the variation of the action takes the form
\begin{equation}
\delta \mathcal{W}=\int d^4x\,\bigl(\delta\mathcal{L}+\partial_\nu\mathcal{L}\,\delta x^\nu\bigr),
\end{equation}
Imposing invariance of the action leads to the local Noether relation as\cite{greiner2013field},
\begin{equation}
\begin{aligned}
0=\;&
\Bigl(S^\mu-\partial_\nu Q^{\nu\mu}-\partial_\nu\Box R^{\nu\mu}\Bigr)\widetilde{\delta}A_\mu
\\[4pt]
&+
\partial_\nu\Big[
Q^{\nu\mu}\widetilde{\delta}A_\mu
+R^{\nu\mu}\Box\widetilde{\delta}A_\mu
-\partial_\rho R^{\nu\mu}\partial^\rho\widetilde{\delta}A_\mu\\[4pt]
&\qquad \qquad \qquad \qquad \qquad+\Box R^{\nu\mu}\widetilde{\delta}A_\mu
+\mathcal{L}\delta x^\nu
\Big].
\end{aligned}
\label{eq:Noether-master1}
\end{equation}
This is an {equation of continuity} for the gauge field defined by\cite{greiner2013field},
\begin{equation}
\partial_\nu \mathscr{F}^\nu(x) = 0,
\end{equation}
with the RGUP-modified current density
\begin{align}
\mathscr{F}^\nu(x)
&=
Q^{\nu\mu}\widetilde{\delta}A_\mu
+R^{\nu\mu}\Box\widetilde{\delta}A_\mu
-\partial_\rho R^{\nu\mu}\partial^\rho\widetilde{\delta}A_\mu
\nonumber \\[2pt]
& \qquad \qquad \qquad\qquad+\Box R^{\nu\mu}\widetilde{\delta}A_\mu
+\mathcal{L}\delta x^\nu.
\label{eq:Noether-master}
\end{align}
Using this conserved current, one can extract the physically relevant Noether charges associated with spacetime symmetries of the RGUP-modified action.  Invariance under infinitesimal spacetime translations leads to the definition of the canonical energy momentum tensor, whose components describe the local densities and fluxes of energy and momentum.  Also, invariance under infinitesimal Lorentz transformations gives rise to the conserved angular momentum tensor, whose spatial components determine the angular-momentum density of the field.  This tensor naturally separates into orbital and spin contributions, providing the basis for the analysis of orbital angular momentum and intrinsic spin in the RGUP-modified gauge theory. 

\textbf{{Invariance Under Spacetime Translation}}:
For spacetime translations the shape of the field supposed not to change that is $\Delta A_{\mu}=0$ so, $\widetilde{\delta}A_\mu=-\delta x^\lambda\partial_\lambda A_\mu$ and\eqref{eq:Noether-master} reduces to a conservation law.  
Extracting the coefficient of $\delta x^\lambda$ gives the modified canonical energy-momentum tensor
\begin{align}
\mathcal{T}^{\nu}_{\lambda}
&=
Q^{\nu\mu}\partial_{\lambda}A_{\mu}
+
R^{\nu\mu}\partial_{\rho}\partial^{\rho}\partial_{\lambda}A_{\mu}
-
(\partial_\rho R^{\nu\mu})\partial^{\rho}\partial_{\lambda}A_{\mu}\nonumber\\[3pt]
&\qquad \qquad \qquad \qquad+
(\Box R^{\nu\mu})\partial_{\lambda}A_{\mu}
-
\mathcal{L}\delta^{\nu}_{\lambda}.
\end{align}
Using Eq.(\ref{emt}), stress energy momentum tensor express as,
\begin{equation}\label{EMTF}
\begin{aligned}
\mathcal{T}^{\nu}_{\lambda}
=&\;
-F^{\nu\mu}\partial_{\lambda}A_{\mu}
-2\gamma\,\Box F^{\nu\mu}\partial_{\lambda}A_{\mu}
-\gamma F^{\nu\mu}\partial_{\rho}\partial^\rho\partial_{\lambda}A_{\mu}
\\[3pt]
&\qquad \qquad \qquad \qquad \;
+\gamma(\partial_{\rho}F^{\nu\mu})\partial^{\rho}\partial_{\lambda}A_{\mu}
-\mathcal{L}\delta^{\nu}_{\lambda},
\end{aligned}
\end{equation}
which is the canonical energy-momentum tensor of the RGUP-modified gauge field.
The RGUP-modified canonical energy-momentum tensor exhibits a clear structural
decomposition into the standard Maxwell contribution and deformation-induced
corrections depend on the parameter $\gamma$. In the limit
$\gamma\rightarrow 0$, the tensor reduces smoothly to the familiar Maxwell form \cite{lorce2015light},
\begin{equation*}
T^{\nu}_{\lambda}\big|_{\gamma=0}
=
- F^{\nu\mu}\,\partial_{\lambda}A_{\mu}
- \mathcal{L}\,\delta^{\nu}_{\lambda},
\end{equation*}
while the RGUP corrections arise entirely from higher-derivative terms present
in the effective Lagrangian. In particular, the appearance of the
d'Alembertian operator acting on the field strength, $\Box F_{\mu\nu}$,
generates contributions involving up to third-order derivatives of the gauge
potential $A_\mu$, a generic hallmark of quantum-gravity-inspired effective
field theories encoding nonlocality and minimal-length effects. Also it is expected
from a Noether construction, the resulting tensor is canonical in nature and
therefore, neither manifestly symmetric nor gauge invariant at the operator
level. The lack of gauge invariance of the canonical energy-momentum tensor is not merely a
formal issue.  In a gauge theory the fields contain unphysical degrees of freedom, and
physical quantities must therefore be invariant under gauge transformations \cite{leader2011controversy}.  A
non-gauge-invariant energy-momentum tensor assigns different energy or momentum
densities to physically equivalent configurations, making it unsuitable as a physical
observable.  This motivates the use of a symmetric and gauge-invariant (Belinfante or
kinetic) energy-momentum tensor to define physically meaningful energy, momentum, and
angular momentum densities\cite{leader2011controversy,belinfante1940current}.
 The inclusion of RGUP corrections
preserves this structural property while enriching the energy-momentum tensor
with higher-derivative contributions that modify local energy and momentum
densities.

\textbf{{Under Lorentz Invariance:}}
\label{sec:Lorentz Invariance}

Spacetime is assumed to be homogeneous and isotropic, so that the physical laws are independent of position and orientation. This implies that the action of the theory must be invariant under Lorentz transformations. These transformations include ordinary spatial rotations as well as Lorentz boosts, which mix space and time coordinates.

An infinitesimal Lorentz transformation can be written as,
\begin{equation}\label{transformationlo}
x'^{\mu} = x^{\mu} + \delta\omega^{\mu\nu} x_{\nu},
\end{equation}
where the parameters $\delta\omega^{\mu\nu}$ are antisymmetric, $\delta\omega^{\mu\nu} = -\delta\omega^{\nu\mu}$, and represent an infinitesimal rotation or boost in spacetime.

For infinitesimal Lorentz transformations the RGUP modified Noether current is proportional to the angular-momentum tensor, and can be written as the contraction of the antisymmetric Lorentz parameter with a rank-three tensor \(M_{\lambda}{}_{\mu\nu}\)\cite{greiner2013field}.  This tensor is defined as the sum of an orbital contribution built from the RGUP modified canonical energy-momentum tensor and an intrinsic spin contribution associated with the transformation of the gauge field. The conservation of the RGUP modified Noether current therefore implies the local conservation of the angular-momentum tensor, which provides the fundamental densities used to analyze orbital and spin angular momentum in the RGUP-modified theory.
The gauge field $A_{\mu}$ transforms linearly under the same symmetry. The infinitesimal variation of the field is defined as \cite{greiner2013field,leader2011controversy},
\begin{equation}\label{ause}
\Delta A_{\mu}(x)
\equiv A'_{\mu}(x') - A_{\mu}(x)
= \frac{1}{2}\,\delta\omega^{\rho\sigma}(I_{\rho\sigma})_{\mu}^{\nu} A_{\nu}(x),
\end{equation}
where $(I_{\rho\sigma})_{\mu}^{\nu}$ are the generators of the Lorentz group in the vector representation, explicitly given by\cite{leader2011controversy},
\begin{equation}
(I_{\rho\sigma})_{\mu}^{\nu}
= i(g_{\rho}^{\nu}\delta_{\sigma\mu} - g_{\sigma}^{\nu}\delta_{\rho\mu}).
\end{equation}

These generators encode the infinitesimal mixing of the field components under Lorentz rotations and boosts, and their anti-symmetry reflects the anti-symmetry of the Lorentz transformation parameters. This transformation law follows directly from the requirement that $A_{\mu}$ transforms as a Lorentz four-vector and provides the starting point for the construction of the conserved Lorentz (angular momentum) current via Noether’s theorem \cite{halder2018noether}. Now, substituting the transformation equations Eqs.~(\ref{transformationlo}, \ref{ause})in the  the functional variation defination of  Eq.~(\ref{eq:tilde-delta}) give \begin{align}
\widetilde{\delta}A_\mu
&= \frac{1}{2}\delta\omega^{\alpha\beta}(I_{\alpha\beta})_\mu^\nu A_\nu
- \delta\omega^{\alpha\beta}x_\beta\partial_\alpha A_\mu.
\end{align}
Using the higher-order Euler–Lagrange equation,
\begin{equation}
S^\mu-\partial_\nu Q^{\nu\mu}-\partial_\nu\Box R^{\nu\mu}=0,
\end{equation}
 and the RGUP-modified Noether current density defined in Eq.~(\ref{eq:Noether-master1}), will provide a total divergence
\begin{equation}
\partial_\nu \mathscr{F}^\nu(x)=0.
\end{equation}
The current density $\mathscr{F}^\nu$ is computed as
\begin{equation}
\mathscr{F}^\nu(x)
= \frac{1}{2}\delta\omega^{\alpha\beta} J^{\nu}_{\alpha\beta,\small{\text{RGUP}}}(x),
\end{equation}
where
\begin{align}
J^{\nu}_{\alpha\beta,\mathrm{RGUP}}
=&\;
\left(
\mathcal{T}^{\nu}_{\alpha} x_{\beta}
-
\mathcal{T}^{\nu}_{\beta} x_{\alpha}
\right)-
\Bigl[
Q^{\nu\mu}
+
R^{\nu\mu}\partial^{\rho}\partial_{\rho}
\nonumber\\
&\qquad \qquad 
-
(\partial_{\rho}R^{\nu\mu})\partial^{\rho}
+
\Box R^{\nu\mu}
\Bigr]
\left(I_{\alpha\beta}\right)^\nu_\mu A_{\nu}.
\end{align}
The conservation law $\partial_\nu \mathscr{F}^\nu=0$ therefore implies
\begin{equation}
\partial_\nu J^{\nu}_{\alpha\beta,\small{\text{RGUP}}}=0,
\end{equation}
which expresses the local conservation of the total angular momentum in the RGUP-modified gauge theory.

\subsection*{Angular Momentum of the Gauge Field with RGUP Corrections}
After establishing the construction of the modified-EMT in the above section, we are now in a position to construct the associated angular momentum of the RGUP-modified gauge field. 
In field theory, angular momentum arises as the Noether charge corresponding to Lorentz invariance, 
and is constructed from the canonical energy-momentum tensor together with the intrinsic Lorentz transformation properties of the fields.

Since the tensor in Eq.(\ref{EMTF}) is canonical rather than symmetric, 
the resulting angular momentum naturally decomposes into orbital and spin contributions. 
This decomposition is present already in ordinary Maxwell theory, however, 
the RGUP-induced higher-derivative terms enrich both sectors by modifying the local densities while preserving the total conserved charge.

The standard canonical angular momentum density decomposes into orbital and spin density terms 
\begin{equation}
J^{\mu \nu \alpha}= M^{\mu \nu \alpha}_{OAM}+ S^{\mu \nu \alpha}_{spin},\quad \text{with} \qquad \partial_\mu J^{\mu\nu \alpha}=0,
\end{equation}
where the standard OAM part is defined as
\begin{equation}\label{orbitalcanonical}
M^{\mu \nu \alpha}_\text{OAM}=x^\nu T^{\mu \alpha}- x^\alpha T^{\mu \nu }
\end{equation} and the spin term  is defined by the intrinsic Lorentz transformation of the gauge field $A_\mu$,
\begin{equation}\label{spincanonical}
S^{\mu\nu\alpha}_{\text{spin}} = \frac{\partial \mathcal{L}}{\partial(\partial_\mu A_\beta)}
\Big( \eta^{\nu\beta} A^\alpha - \eta^{\alpha\beta} A^\nu \Big),  
\end{equation}
We now analyze the orbital and spin part contribution to the gauge field within the RGUP frame work which follows directly from the canoncial EMT. Ths allows us to explicitly track the modification induced by RGUP corrections at the level of local densities. 
%In a gauge Field, the total angular momentum of the field arises from the canonical Noether current.  With the RGUP-modified Lagrangian, both the orbital and spin parts acquire additional higher-derivative contributions.  This subsection summarizes the full structure of the angular momentum density and its integrated vector form, which together describe the complete angular momentum carried by the gauge field when RGUP corrections are present.

\subsubsection*{Modified Orbital Angular Momentum}

In the canonical formulation, the orbital angular momentum (OAM) current is constructed the energy-momentum tensor and the spacetime coordinates given in Eq.(\ref{orbitalcanonical}).
Since physical angular momentum is associated with spatial rotations, we focus on the spatial components with one temporal index, $(0jk)$, for which the orbital angular momentum current takes the form 
\begin{equation}\label{used1}
M_{\text{OAM},\text{RGUP}}^{0jk}
=
x^{j}\mathcal{T}^{0k}
-
x^{k}\mathcal{T}^{0j}.
\end{equation}
Within the RGUP framework, all Planck-scale corrections enter exclusively through the modified EMT, while the spacetime coordinates themselves remain undeformed.On substituting the expression given in Eq.~(\ref{EMTF}) and using $F^{0l}=-E^{l}$ to Eq.(\ref{used1}) yields the RGUP-corrected orbital angular momentum density
\begin{equation}
\begin{aligned}
M_{\text{OAM,RGUP}}^{0jk}
&=
x^{j}E^{l}\partial^{k}A_{l}
-
x^{k}E^{l}\partial^{j}A_{l}
\\
&\quad
+2\gamma\Bigl[
x^{j}(\Box E^{l})\partial^{k}A_{l}
-
x^{k}(\Box E^{l})\partial^{j}A_{l}
\Bigr]
\\
&\quad
+\gamma\Bigl[
x^{j}E^{l}\Box\partial^{k}A_{l}
-
x^{k}E^{l}\Box\partial^{j}A_{l}
\Bigr]
\\
&\quad
-\gamma\Bigl[
x^{j}(\partial_{\rho}E^{l})\partial^{\rho}\partial^{k}A_{l}
-
x^{k}(\partial_{\rho}E^{l})\partial^{\rho}\partial^{j}A_{l}
\Bigr].
\end{aligned}
\end{equation}

It is convenient to express the OAM density in vector form by contracting the antisymmetric spatial indices using the Levi-Civita tensor $\epsilon^{ijk}$,
%as\begin{equation}L^{i}(x)=\frac{1}{2}\epsilon^{ijk}M_{\text{OAM,RGUP}}^{0jk},\end{equation}
which simplifies to
\begin{align}
L^{i}(x)
&=
E^{l}(\mathbf{x}\times\nabla)^{i}A_{l}
+\gamma\Bigl[
2(\Box E^{l})(\mathbf{x}\times\nabla)^{i}A_{l}
\nonumber \\
&\qquad\quad
+
E^{l}(\mathbf{x}\times\nabla)^{i}(\Box A_{l})
-
(\partial_{\rho}E^{l})(\mathbf{x}\times\nabla)^{i}(\partial^{\rho}A_{l})
\Bigr].
\end{align}
The integrated orbital angular momentum vector, representing the physical OAM carried by the gauge field under RGUP framework, is therefore given by
\begin{align}
\mathbf{L}
&=
\int d^{3}x\,
\Bigl[
E^{l}(\mathbf{x}\times\nabla)A_{l}
+\gamma\Bigl\{
2(\Box E^{l})(\mathbf{x}\times\nabla)A_{l}
\nonumber \\
&\qquad\quad
+
E^{l}(\mathbf{x}\times\nabla)(\Box A_{l})
-
(\partial_{\rho}E^{l})(\mathbf{x}\times\nabla)(\partial^{\rho}A_{l})
\Bigr\}
\Bigr].
\end{align}

From the above expression, one observes that the corrections appear explicitly in the orbital angular momentum density, modifying its local structure through the presence of higher-derivative terms. Nevertheless, the orbital angular momentum continues to transform as a three-vector under spatial rotations, thereby preserving its overall vectorial character. The leading term reproduces the standard Maxwell result, ensuring a smooth correspondence with the classical limit. In the limit $\gamma \to 0$, one recovers
\begin{equation}
\mathbf{L}_{0}
=
\int d^{3}x\,E^{l}(\mathbf{x}\times\nabla)A_{l}.
\end{equation}

\subsubsection*{Spin Angular Momentum}
The spin part is obtained from the internal Lorentz transformation of the vector field. The RGUP-modified spin current is defined as
\begin{equation}
    S^{\mu \nu \lambda}_{spin, RGUP}=\Pi^{\lambda \alpha}_{RGUP}(\Sigma^{\mu \nu})^{\beta}_{\alpha}A_{\beta}
\end{equation}
where, the minimal length corrections are present in $\Pi^{\lambda \alpha}_{RGUP}=-F^{\lambda \alpha}-2\gamma\Box F^{\lambda \alpha}-\gamma F^{\lambda \alpha}\Box +\gamma(\partial_{\rho} F^{\lambda \alpha})\partial^{\rho}$ and $(\Sigma^{\mu \nu})^{\beta}_{\alpha}=i\left(\delta^{\mu}_{\alpha} g^{\nu\beta}-\delta^{\nu}_{\alpha} g^{\mu\beta}\right)
$ is the spin-1 lorentz generator\cite{leader2011controversy}. Using $\Pi^{0j}=-E^{j}$ we obtain;
\begin{equation}
\Pi_{\text{RGUP}}^{0j}
=
E^{j}
+2\gamma\,\Box E^{j}
+\gamma E^{j}\Box
-\gamma(\partial_{\rho}E^{j})\partial^{\rho}
\end{equation}
the spin tensor becomes
\begin{equation}
S_{\text{spin,RGUP}}^{0jk}
=
\Pi_{\text{RGUP}}^{0j}A^{k}-\Pi_{\text{RGUP}}^{0k}A^{j}.
\end{equation}

The corresponding spin vector density is
\begin{align}
S^{i}(x)
=
&(\mathbf{E}\times\mathbf{A})^{i}
+ 2\gamma[(\Box\mathbf{E})\times\mathbf{A}]^{i}
\nonumber\\
&+ \gamma[\mathbf{E}\times(\Box\mathbf{A})]^{i}
- \gamma[(\partial_{\rho}\mathbf{E})\times(\partial^{\rho}\mathbf{A})]^{i}.
\end{align}
Thus, the integrated spin angular momentum vector of the gauge field is
\begin{equation}
\begin{aligned}
\mathbf{S}
=
\int d^{3}x\,
\Bigl[
&\mathbf{E}\times\mathbf{A}
+2\gamma(\Box\mathbf{E})\times\mathbf{A}
\\
&+\gamma\,\mathbf{E}\times(\Box\mathbf{A})
-\gamma(\partial_{\rho}\mathbf{E})\times(\partial^{\rho}\mathbf{A})
\Bigr].
\end{aligned}
\end{equation}
The standard spin vector  is recovered for $\gamma\to 0$:
\begin{equation}
\mathbf{S}_{0}
=
\int d^{3}x\,\mathbf{E}\times\mathbf{A}.
\end{equation}

\subsubsection*{Total gauge Angular Momentum with RGUP}

The RGUP-modified total angular momentum of the gauge field is expressed as the sum of its orbital and spin parts
\begin{equation}
\mathbf{J}
=
\mathbf{L}
+
\mathbf{S}.
\end{equation}
RGUP corrections modify both contributions through higher-derivative couplings involving 
$\Box E^{l}$, $\Box A_{l}$, $\partial_{\rho}E^{l}$ and $\partial_{\rho}A_{l}$,  
showing that the generalized uncertainty principle alters the internal structure of electromagnetic angular momentum while maintaining a smooth limit to Maxwell theory.\\

\textbf{{Conserved spin and  OAM currents}}:
Noether's theorem establishes a direct connection between continuous
symmetries of the action and conserved quantities.  In particular, invariance of the
Lagrangian under space-time translations implies conservation of energy and momentum. As discussed
in the previous section, even in the presence of RGUP-induced modifications, the
theory remains invariant under infinitesimal spacetime translations. Consequently,
the RGUP-modified canonical EMT,
$\mathcal{T}^{\nu}_{\lambda}$, satisfies a local conservation law,
\begin{equation}
\partial_\nu \mathcal{T}^{\nu}{}_{\lambda} = 0,
\qquad
\mathcal{T}^{\nu}{}_{\lambda} \neq \mathcal{T}_{\lambda}{}^{\nu},
\label{eq:EMT_conservation_nonsymmetry}
\end{equation}

It is important to emphasize that the modified canonical energy-momentum tensor is
not symmetric
In addition to translation invariance, invariance of the action under Lorentz
transformations implies conservation of total angular momentum.  The corresponding
Noether current is a rank-three tensor $J^{\mu\nu\alpha}$, antisymmetric in the Lorentz
indices $\mu$ and $\nu$, which represents the relativistic angular-momentum density of
the field.
The angular-momentum tensor obeys the local conservation law,
\begin{equation}
\partial_\alpha J^{\mu\nu\alpha}_{\mathrm{RGUP}} = 0,
\qquad
J^{\mu\nu\alpha}_{\mathrm{RGUP}}
=
-\,J^{\nu\mu\alpha}_{\mathrm{RGUP}},
\label{eq:RGUP_AM_conservation_antisymmetry}
\end{equation}

and is antisymmetric in its first two indices,

The canonical modified orbital angular momentum density is defined as
\begin{equation}
M^{\nu}_{\alpha\beta,RGUP}
=
\mathcal{T}^{\nu}_{\alpha} x_{\beta}
-
\mathcal{T}^{\nu}_{\beta} x_{\alpha},
\end{equation}
On taking the divergence and using standard product-rule identities, one finds
\begin{equation}
\partial_{\nu} M^{\nu}_{\alpha\beta,RGUP}
=
(\partial_{\nu}\mathcal{T}^{\nu}_{\alpha}) x_{\beta}
-
(\partial_{\nu}\mathcal{T}^{\nu}_{\beta}) x_{\alpha}
+
\mathcal{T}_{\beta\alpha}
-
\mathcal{T}_{\alpha\beta}.
\end{equation}
Imposing translational invariance and using \ref{eq:EMT_conservation_nonsymmetry} , this reduces to
\begin{equation}
\partial_{\nu} M^{\nu}_{\alpha\beta,RGUP}
=
\mathcal{T}_{\beta\alpha}
-
\mathcal{T}_{\alpha\beta}.
\end{equation}
Since the RGUP-modified canonical energy-momentum tensor is not symmetric, the right-hand side does not vanish, implying
\begin{equation}
\partial_{\nu} M^{\nu}_{\alpha\beta,RGUP} \neq 0.
\end{equation}
Hence, the modified  canonical orbital angular momentum is not conserved by itself.

The RGUP spin current is expressed as,
\begin{equation}
S^{\mu\nu\lambda}_{\text{spin},RGUP}
=
i\left(
\Pi^{\lambda\mu} A^\nu
-
\Pi^{\lambda\nu} A^\mu
\right),
\end{equation}
with RGUP corrected canonical conjugate momentum
\begin{equation}
\Pi^{\lambda\mu}_{RGUP}
=
-
F^{\lambda\mu}
-
2\gamma \Box F^{\lambda\mu}
+
\gamma (\partial_{\rho}F^{\lambda\mu})\partial^{\rho}.
\end{equation}

Taking the divergence and applying the Leibniz rule,
expanding $\partial_\lambda \Pi^{\lambda\mu}_{RGUP}$ and using only
index symmetry and commutativity of partial derivatives,if $\gamma=0$. The divergence reduces to
\begin{equation}
\partial_\lambda S^{\mu\nu\lambda}_{\text{spin},RGUP}
=
i\left[
A^\mu\,\partial_\lambda F^{\lambda\nu}
-
A^\nu\,\partial_\lambda F^{\lambda\mu}
\right].
\end{equation}

Since the RGUP-modified theory preserves the standard field-theoretic
structure of the spin density, the spin current does not satisfy an
independent conservation law. Consequently, the RGUP-modified spin density
is also not conserved on its own.
Thus, the spin current is not conserved in general,
\begin{equation}
\partial_\lambda S^{\mu\nu\lambda}_{\text{spin},RGUP} \neq 0,
\end{equation}

These two parts are not conserved separately.  
\begin{equation}
\partial_\lambda S^{\mu\nu\lambda}_{spin,RGUP} \neq 0,
\qquad
\partial_\nu M^{\nu}_{\alpha\beta,RGUP} \neq 0,
\end{equation}
which shows explicitly that both modified OAM current and Spin current  are nonzero.
 Although RGUP corrections modify the explicit forms of the orbital and spin
angular momentum densities, the standard Noether-theoretic conclusion remains
unchanged orbital and spin angular momenta are not conserved separately.
Instead, RGUP effects redistribute angular momentum between these sectors,
while the total angular momentum current remains conserved. This demonstrates
that the non-separate conservation of spin and orbital angular momentum is
retained within the RGUP framework, and constitutes a distinctive result of
our analysis.

\section{Symmetric Energy-Momentum Tensor for the RGUP-Modified Gauge Field}\label{sec4}

In gauge theories the energy-momentum tensor obtained from Noether's first theorem
is generally not symmetric and is not gauge invariant \cite{leader2011controversy}. This difficulty appears in
both Abelian and non-Abelian theories and typically leads to the introduction of 
Belinfante-type improvement terms, which are neither unique nor canonical. 
A consistent and unambiguous construction is instead obtained from the framework of 
Noether's second theorem, which incorporates the full local symmetry of the theory.
As shown in Ref.\cite{freese2022noether}, when the gauge field is allowed to transform
appropriately under local translations, the resulting energy-momentum tensor (EMT)
is automatically symmetric and gauge invariant without the need for additional improvement. 
We follow this framework for the RGUP-modified gauge field Lagrangian given in Eq.(\ref{Lagrangian}). We introduce the generalized variational derivatives as eq; \ref{DERIVATIVE HIGHER ORDER}
which are evaluated as \ref{emt}.

To derive the symmetric EMT, we employed a local translation \cite{freese2022noether},

\begin{equation}\label{tranlocal}
x^\mu \rightarrow x'^\mu = x^\mu + \xi^\mu(x),
\end{equation}
 The transformations of the gauge field and its derivatives under the infinitesimal coordinate shift are given by the following relations (shown in Appendix \ref{appa}),
\begin{align}
\Delta A_\mu
&= -(\partial_\mu \xi^\lambda)\,A_\lambda ,
\\[6pt]
\Delta(\partial_\nu A_\mu)
&= -(\partial_\nu \xi^\lambda)\,\partial_\lambda A_\mu
   -(\partial_\mu \xi^\lambda)\,\partial_\nu A_\lambda ,
\\[6pt]
\Delta(\Box\,\partial_\nu A_\mu)
&= -(\partial_\nu \xi^\lambda)\,\Box(\partial_\lambda A_\mu)
   -(\partial^\rho \xi^\beta)\,\partial_\rho\partial_\beta\partial_\nu A_\mu
\nonumber\\
&\quad
   -(\partial_\rho \xi^\alpha)\,\partial_\alpha\partial^\rho\partial_\nu A_\mu
   -(\partial_\mu \xi^\sigma)\,\partial_\rho\partial^\rho\partial_\nu A_\sigma .
\end{align}

Next, the total variation of the action $\Delta \mathcal{W}= \mathcal{W}'-\mathcal{W}$ in the presence of the transformation Eq.(\ref{tranlocal}),
\begin{equation}
\Delta \mathcal{W} = \int d^{4}x\,
\bigl[
\Delta\mathcal{L} + (\partial_\mu\xi^\mu)\mathcal{L}
\bigr],
\end{equation}
includes the Jacobian factor $\partial_\mu\xi^\mu$ arising from the variation of the
integration measure. On applying the chain rule to the variation of the Lagrangian density give
\begin{equation}
\Delta\mathcal{L}
=
S^\mu\Delta A_\mu
+
Q^{\nu\mu}\Delta(\partial_\nu A_\mu)
+
R^{\nu\mu}\Delta(\Box\partial_\nu A_\mu).
\end{equation}
Both $\Delta A_\mu$ and its derivative variations are linear in derivatives of 
$\xi^\mu(x)$. Thus, one can define a operator as
\begin{align}
\mathscr{D}^\mu_{\ \nu}[A]
\equiv {} &
- \frac{\partial}{\partial(\partial_\mu \xi^\nu)}
\Bigl[
S^\alpha \Delta A_\alpha
+ Q^{\rho\alpha} \Delta(\partial_\rho A_\alpha)
\nonumber\\[4pt]
&\qquad\qquad\qquad\qquad
+ R^{\rho\alpha} \Delta(\Box \partial_\rho A_\alpha)
\Bigr].
\end{align}

Using this definition, the action variation becomes
\begin{equation}
\Delta \mathcal{W}
=
\int d^4x \Bigl[
-\mathscr{D}^\mu_{\ \nu}[A]\,\partial_\mu\xi^\nu
+
(\partial_\mu\xi^\mu)\mathcal{L}
\Bigr],
\end{equation} Integrating the second term by parts and dropping the boundary terms gives
\begin{equation}
\Delta \mathcal{W}
=
\int d^4x\,\xi^\nu(x)\,
\partial_\mu
\Bigl[
\mathscr{D}^\mu_{\ \nu}[A]
-
\delta^\mu_{\ \nu}\mathcal{L}
\Bigr].
\end{equation}
Since the local translation function $\xi^\nu(x)$ is arbitrary apart from compact
support, invariance of the action $(\Delta S=0)$ implies
\begin{equation}
\partial_\mu
\Bigl[
\mathscr{D}^\mu_{\ \nu}[A]
-
\delta^\mu_{\ \nu}\mathcal{L}
\Bigr]
=0 .
\end{equation}
Thus the symmetric, gauge-invariant energy-momentum tensor is
\begin{equation}
\mathscr{T}^\mu_{\ \nu}
=
\mathscr{D}^\mu_{\ \nu}[A]
-
\delta^\mu_{\ \nu}\mathcal{L}.
\end{equation}
After evaluating $\mathscr{D}^\mu_{\ \nu}[A]$ using the explicit RGUP expressions for 
$Q^{\mu\nu}$ and $R^{\mu\nu}$ yields the symmetric EMT as
\begin{align}\label{belin}
\mathscr{T}^\nu_{\lambda}
&=
(-F^{\nu\mu}-\gamma\Box F^{\nu\mu})\,\partial_\lambda A_\mu
+
(-F^{\mu\nu}-\gamma\Box F^{\mu\nu})\,\partial_\mu A_\lambda
\notag\\
&\qquad \quad
- \gamma F^{\mu\nu}\,\Box\partial_{\mu}A_{\lambda}
- \gamma F^{\nu\mu}\,\Box\partial_{\lambda}A_{\mu}
\notag\\
&\qquad \qquad \quad
- 2\gamma F^{\alpha\mu}\,
  \partial_{\lambda}\partial^{\nu}\partial_{\alpha}A_{\mu}
- \delta^\nu_{\lambda}\mathcal{L}.
\end{align}

The resulting tensor is symmetric in its spacetime indices, gauge invariant, and in limit $\gamma\to 0$, the RGUP-induced contributions vanish identically,
and the energy–momentum tensor reduces smoothly to the Belinfante-improved EMT.  
Eq.(\ref{belin}) represents the unique EMT associated with local translation symmetry in the
minimal length modified gauge field theory.

\subsection{Minimal length corrections in the Belinfante Angular Momentum }
To compute the angular momentum carried by the RGUP-modified electromagnetic
field, we employ the symmetric (Belinfante) energy–momentum tensor obtained
from invariance of the action under local spacetime translations \cite{belinfante1940current}. Raising one
index with the Minkowski metric, the Belinfante tensor takes the form
%In order to compute the angular momentum carried by the RGUP-modified Gauge field, we use the symmetric (Belinfante) energy-momentumtensor obtained from local translations.  With one index raised it reads
\begin{align}
\mathscr{T}^{\nu\beta}&=\mathscr{T}^{\nu}_{\lambda}\,\eta^{\lambda\beta}
 \nonumber\\
&=
S^{\nu}A^{\beta}
+Q^{\nu\mu}\,\partial^{\beta}A_{\mu}
+Q^{\mu\nu}\,\partial_{\mu}A^{\beta}
\nonumber\\
&\quad
+R^{\mu\nu}\,\Box\partial_{\mu}A^{\beta}
+R^{\nu\mu}\,\Box\partial^{\beta}A_{\mu}
\nonumber\\
&\quad
+2R^{\alpha\mu}\,\partial^{\beta}\partial^{\nu}\partial_{\alpha}A_{\mu}
-\eta^{\nu\beta}\mathcal{L},
\label{T-raised-Bel}
\end{align}
where the tensors $S^{\nu}$, $Q^{\nu\mu}$ and $R^{\nu\mu}$ encode the
dependence of $\mathcal{L}$ on $A_{\mu}$, $\partial_{\nu}A_{\mu}$ and
$\Box\partial_{\nu}A_{\mu}$, respectively.
For the RGUP-modified electromagnetic Lagrangian one has
Eq.(\ref{emt})
in which all minimal-length effects enter exclusively through
higher-derivative corrections to the field strength.

The Belinfante angular momentum density associated with a symmetric EMT is given as,
\begin{equation}
M^{\nu\gamma\beta}_{\text{RGUP}}
=
x^{\gamma}\mathscr{T}^{\nu\beta}
-
x^{\beta}\mathscr{T}^{\nu\gamma},
\label{M-def-Bel}
\end{equation}
and the physically relevant spatial component is
\begin{equation}
M^{0jk}_{\text{RGUP}}
=
x^{j}\mathscr{T}^{0k}
-
x^{k}\mathscr{T}^{0j}.
\label{M-0jk-Bel}
\end{equation}
The corresponding vector density is defined as
\begin{equation}
J^{i}(x)
=
\frac{1}{2}\,\epsilon^{ijk}M^{0jk}_{\text{RGUP}}(x).
\label{J-def}
\end{equation}

To make the role of the RGUP deformations explicit, it is convenient to express the full Belinfante angular momentum density as a modification of the standard Maxwell result. In the{ Maxwell (Belinfante) limit}, corresponding to $\gamma\to0$, the constitutive tensors reduce to
\begin{equation}
Q^{\mu\nu}=-F^{\mu\nu},
\qquad
R^{\mu\nu}=0,
\end{equation}
so that the angular momentum density is entirely determined by the usual Gauge field strength. Using
%One can write the RGUP corrections in the angular momentum density as modification to the standard result of hte maxwells limit, we wrote the corrections induced by the RGUP corrections as explicite corrections to the standad results In {Maxwell (Belinfante) limit}, where $\gamma\to 0$ we have\begin{equation}Q^{\mu\nu}=-F^{\mu\nu},\qquadR^{\mu\nu}=0,\end{equation}and with
\begin{equation}
F^{0a}=-E^{a},
\qquad
F^{a0}=E^{a},
\qquad
F^{k}_{a}=\partial^{k}A_{a}-\partial_{a}A^{k},
\end{equation}
the $0jk$-component of the Belinfante tensor takes the form
\begin{equation}
M^{0jk}
=
x^{j}E^{a}F^{k}_{a}
-
x^{k}E^{a}F^{j}_{a}.
\label{M0jk-Maxwell-Bel}
\end{equation}
Substituting this expression into \eqref{J-def}, one obtains
\begin{equation}
\begin{aligned}
J^{i}_{0}(x)
&=
\frac{1}{2}\epsilon^{ijk}
\bigl(
x^{j}E^{a}F^{k}_{a}
-
x^{k}E^{a}F^{j}_{a}
\bigr)
\\
&=
\epsilon^{ijk}x^{j}E^{a}F^{k}_{a},
\end{aligned}
\end{equation}
where in the second line we have exchanged $j\leftrightarrow k$ in the second
term and used the anti-symmetry of $\epsilon^{ijk}$.
Writing the spatial field-strength components in terms of the magnetic field as
$
F^{k}_{a}=\epsilon^{kma}B^{m}$,
and using the identity of Levi-Civita tensors
\begin{comment}\begin{equation}
\epsilon^{ijk}\epsilon^{kma}
=
\delta^{im}\delta^{ja}
-
\delta^{ia}\delta^{jm},
\end{equation}
\end{comment}
one arrives at the familiar vector form of the Maxwell angular momentum density
\begin{equation}
\mathbf{J}_{0}(x)
=
-\mathbf{x}\times(\mathbf{E}\times\mathbf{B}),
\end{equation}
so that the total angular momentum is given by
\begin{equation}
\mathbf{J}_{0}
=
\int d^{3}x\,\mathbf{J}_{0}(x)
=
-\int d^{3}x\,\mathbf{x}\times(\mathbf{E}\times\mathbf{B}).
\end{equation}
 In the presence of the RGUP deformation parameter $\gamma$, the tensors
$Q^{\mu\nu}$ and $R^{\mu\nu}$ acquire higher-derivative contributions, which in
turn induce corrections to the Belinfante angular momentum density. It is
convenient to decompose the RGUP-modified tensor as
\begin{equation}
M^{0jk}_{\mathrm{RGUP}}
=
M^{0jk}_{0}
+
M^{0jk}_{1}
+
M^{0jk}_{2}
+
M^{0jk}_{3},
\end{equation}
where $M^{0jk}_{0}$ is the Maxwell contribution given in
Eq.~\eqref{M0jk-Maxwell-Bel}, while $M^{0jk}_{1}$, $M^{0jk}_{2}$, and
$M^{0jk}_{3}$ denote the three distinct RGUP-induced correction terms.

The{ first RGUP-contribution} originates from the $\gamma \Box F^{\mu\nu}$ term in
$Q^{\mu\nu}$ and is given by
\begin{equation}
M^{0jk}_{1}
=
\gamma\Bigl[
x^{j}(\Box E^{a})F^{k}_{a}
-
x^{k}(\Box E^{a})F^{j}_{a}
\Bigr].
\end{equation}
Using the definition \eqref{J-def}, the corresponding vector density reads
\begin{equation}
\begin{aligned}
J^{i}_{1}(x)
&=
\frac{1}{2}\epsilon^{ijk}M^{0jk}_{1}
=
\gamma\,\epsilon^{ijk}x^{j}(\Box E^{a})F^{k}_{a},
\end{aligned}
\end{equation}
where, as in the Maxwell case, the antisymmetry of $\epsilon^{ijk}$ renders the
two terms identical. Writing $F^{k}_{a}=\epsilon^{kma}B^{m}$, this expression
can be cast in the compact vector form
\begin{equation}
\mathbf{J}_{1}(x)
=
-\gamma\,\mathbf{x}\times\bigl((\Box\mathbf{E})\times\mathbf{B}\bigr).
\end{equation}

The {second RGUP contribution} arises from the $R^{\mu\nu}=-\gamma F^{\mu\nu}$ term
in the energy-momentum tensor and yields
\begin{equation}
M^{0jk}_{2}
=
\gamma\Bigl[
x^{j}E^{a}(\Box F^{k}_{a})
-
x^{k}E^{a}(\Box F^{j}_{a})
\Bigr].
\end{equation}
The associated angular momentum density is
\begin{equation}
\begin{aligned}
J^{i}_{2}(x)
&=
\frac{1}{2}\epsilon^{ijk}M^{0jk}_{2}
=
\gamma\,\epsilon^{ijk}x^{j}E^{a}(\Box F^{k}_{a}),
\end{aligned}
\end{equation}
which, upon using $F^{k}_{a}=\epsilon^{kma}B^{m}$, becomes
\begin{equation}
\mathbf{J}_{2}(x)
=
-\gamma\,\mathbf{x}\times\bigl(\mathbf{E}\times(\Box\mathbf{B})\bigr).
\end{equation}

The {third RGUP contribution }originates from the  $R^{\alpha\mu}$ term in the
Belinfante energy-momentum tensor \eqref{T-raised-Bel} involving three derivatives acting on the
gauge field. The corresponding angular momentum tensor reads
\begin{equation}
M^{0jk}_{3}
=
2x^{j}R^{\alpha\mu}\,\partial^{k}\partial^{0}\partial_{\alpha}A_{\mu}
-
2x^{k}R^{\alpha\mu}\,\partial^{j}\partial^{0}\partial_{\alpha}A_{\mu},
\end{equation}
with
\begin{equation}
R^{\alpha\mu}=-\gamma F^{\alpha\mu}.
\end{equation}
From the definition \eqref{J-def}, the associated vector density becomes
\begin{equation}
\begin{aligned}
J^{i}_{3}(x)
&=
\frac{1}{2}\epsilon^{ijk}M^{0jk}_{3}
=
-2\gamma\,\epsilon^{ijk}x^{j}F^{\alpha\mu}\,
\partial^{k}\partial^{0}\partial_{\alpha}A_{\mu},
\end{aligned}
\end{equation}
where antisymmetry of $\epsilon^{ijk}$ has again been used to combine the two
terms. Employing the standard $3+1$ decomposition of the field strength,
\begin{equation}
F^{0a}=-E^{a},
\qquad
F^{a0}=E^{a},
\qquad
F^{ab}=-\epsilon^{abm}B^{m},
\end{equation}
and introducing the orbital operator
\begin{equation}
(\mathbf{x}\times\nabla)^{i}
=
\epsilon^{ijk}x^{j}\partial^{k},
\end{equation}
the third RGUP contribution can be written in compact vector form as
\begin{align}
\mathbf{J}_{3}(x)
&=
2\gamma\,\mathbf{x}\times
\Bigl[
E^{a}\,\partial^{0}\partial_{0}\nabla A_{a}
-
E^{a}\,\partial^{0}\partial_{a}\nabla A_{c}
\nonumber\\
&\qquad\qquad\qquad\qquad+
F^{ba}\,\partial^{0}\partial_{a}\nabla A_{b}
\Bigr].
\end{align}
Collecting the standard Maxwell contribution together with all
RGUP-induced corrections, the total Belinfante angular-momentum density of the
RGUP-modified Gauge field can be written in compact form as
\begin{equation}
\mathbf{J}_{\mathrm{Bel}}(x)
=
\mathbf{J}_{0}(x)
+
\mathbf{J}_{1}(x)
+
\mathbf{J}_{2}(x)
+
\mathbf{J}_{3}(x),
\end{equation}
where $\mathbf{J}_{0}$ denotes the usual Maxwell term and
$\mathbf{J}_{1,2,3}$ encode the minimal-length corrections arising from the
RGUP deformation of the gauge-field dynamics.  Explicitly, one finds
\begin{equation}
\begin{aligned}
\mathbf{J}_{\mathrm{Bel}}(x)
&=
-\mathbf{x}\times(\mathbf{E}\times\mathbf{B})
-\gamma\,\mathbf{x}\times\bigl((\Box\mathbf{E})\times\mathbf{B}\bigr)
\\[2pt]
&\quad-\gamma\,\mathbf{x}\times\bigl(\mathbf{E}\times(\Box\mathbf{B})\bigr)
+2\gamma\,\mathbf{x}\times
\Bigl[
E^{a}\,\partial^{0}\partial_{0}\nabla A_{a}
\\[2pt]
&\quad-
E^{a}\,\partial^{0}\partial_{a}\nabla A_{c}
+
F^{ba}\,\partial^{0}\partial_{a}\nabla A_{b}
\Bigr].
\end{aligned}
\end{equation}

The first term reproduces the familiar Belinfante angular-momentum density of
Maxwell electrodynamics, expressed entirely in terms of the gauge-invariant
Poynting vector $\mathbf{E}\times\mathbf{B}$.  The remaining terms are genuine
RGUP corrections proportional to the deformation parameter $\gamma$ and involve
higher derivatives of the Gauge field.  These contributions reflect
the modified phase-space structure induced by the relativistic generalized
uncertainty principle and encode minimal-length effects in the local flow of
angular momentum.

Despite the presence of higher-derivative terms, the Belinfante angular-momentum
density remains gauge invariant and transforms covariantly under Lorentz
transformations.  The RGUP corrections do not introduce any additional surface
terms beyond those already present in the Maxwell theory, ensuring that the
global angular-momentum generator is well defined.

The total Belinfante angular momentum carried by the RGUP-modified
Gauge field is obtained by spatial integration,
\begin{equation}
\mathbf{J}_{\mathrm{Bel}}
=
\int d^{3}x\,\mathbf{J}_{\mathrm{Bel}}(x),
\end{equation}
which smoothly reduces to the standard Maxwell result in the limit
$\gamma\to0$.  This continuity confirms that the RGUP framework provides a consistent deformation of gauge angular momentum, preserving
the conventional theory as its low-energy limit while incorporating
Planck-scale corrections at the level of local angular-momentum density.

\begin{comment}
Collecting the Maxwell term and all RGUP-induced contributions, the full
Belinfante angular momentum density of the RGUP-modified Gauge field
is given by
\begin{equation}
\mathbf{J}_{\mathrm{Bel}}(x)
=
\mathbf{J}_{0}(x)
+
\mathbf{J}_{1}(x)
+
\mathbf{J}_{2}(x)
+
\mathbf{J}_{3}(x),
\end{equation}
or explicitly,
\begin{equation}
\begin{aligned}
\mathbf{J}_{\mathrm{Bel}}(x)
&=
-\mathbf{x}\times(\mathbf{E}\times\mathbf{B})
-\gamma\,\mathbf{x}\times\bigl((\Box\mathbf{E})\times\mathbf{B}\bigr)
\\[2pt]
&\quad-\gamma\,\mathbf{x}\times\bigl(\mathbf{E}\times(\Box\mathbf{B})\bigr)
+2\gamma\,\mathbf{x}\times
\Bigl[
E^{a}\,\partial^{0}\partial_{0}\nabla A_{a}
\\[2pt]
&\quad-
E^{a}\,\partial^{0}\partial_{a}\nabla A_{0}
+
\epsilon^{abm}B^{m}\,\partial^{0}\partial_{a}\nabla A_{b}
\Bigr].
\end{aligned}
\end{equation}
The total Belinfante angular momentum carried by the RGUP-modified
Gauge field is finally obtained as
\begin{equation}
\mathbf{J}_{\mathrm{Bel}}
=
\int d^{3}x\,\mathbf{J}_{\mathrm{Bel}}(x),
\end{equation}
which continuously reduces to the standard Maxwell Belinfante result in the
limit $\gamma\to0$.
\end{comment}

\subsection{Conservation of the Belinfante Angular-Momentum Current}

We define the angular-momentum current constructed from the symmetric
(Belinfante) energy-momentum tensor as\cite{bliokh2014conservation,leader2011controversy},
\begin{equation}
M^{\nu\gamma\beta}(x)
=
x^{\gamma} \mathscr{T}^{\nu\beta}(x)
-
x^{\beta} \mathscr{T}^{\nu\gamma}(x).
\end{equation}

Taking the divergence and applying the product rule, we obtain
\begin{align}
\partial_\nu M^{\nu\gamma\beta}
&=
\partial_\nu \left( x^{\gamma} \mathscr{T}^{\nu\beta} - x^{\beta} \mathscr{T}^{\nu\gamma} \right)
\end{align}

Using the identity $\partial_\nu x^\gamma = \delta^\gamma_\nu$, this becomes
\begin{align}
\partial_\nu M^{\nu\gamma\beta}
&=
\mathscr{T}^{\gamma\beta}
-
\mathscr{T}^{\beta\gamma}
+
x^\gamma \partial_\nu \mathscr{T}^{\nu\beta}
-
x^\beta \partial_\nu \mathscr{T}^{\nu\gamma}.
\end{align}

For the Belinfante energy-momentum tensor the symmetry and conservation
conditions
\begin{equation}
\mathscr{T}^{\gamma\beta} = \mathscr{T}^{\beta\gamma},
\qquad
\partial_\nu \mathscr{T}^{\nu\beta} = 0
\end{equation}
hold. Substituting these into the previous expression yields
\begin{equation}
\partial_\nu M^{\nu\gamma\beta} = 0.
\end{equation}

Therefore, the angular-momentum current constructed from the Belinfante
energy-momentum tensor is locally conserved.

\section{APPLICATION OF THE SYMMETRIC ANGULAR MOMENTUM: RGUP-Modified Poynting Vector}\label{sec5}

In any field theory with a symmetric energy-momentum tensor,the local density of angular momentum is fully determined by the momentum
density $T^{0i}$. We start with the RGUP-modified definition of the angular-momentum
current,
\begin{equation}
M^{\lambda\mu\nu}(x)
=
x^{\mu} \mathscr{T}^{\lambda\nu}(x)
-
x^{\nu} \mathscr{T}^{\lambda\mu}(x),
\label{eq:Mlam}
\end{equation}
the spatial angular-momentum density follows by setting $\lambda=0$,
\begin{equation}
J^{i}(x)
=
\frac{1}{2}\,\epsilon^{ijk} M^{0jk}(x)
=
\epsilon^{ijk} x^{j} \mathscr{T}^{0k}(x).
\label{eq:J-from-T0i}
\end{equation}
Equation~\eqref{eq:J-from-T0i} makes explicit that the quantity $\mathscr{T}^{0k}$ plays the
role of the local momentum density carried by the field. It is therefore natural
to introduce a vector field $\mathbf{S}(x)$ defined through
\begin{equation}
S^{k}(x)=\mathscr{T}^{0k}(x),
\qquad\Rightarrow\qquad
\mathbf{J}(x)=\mathbf{x}\times\mathbf{S}(x).
\label{eq:S-def}
\end{equation}
which generalizes the notion of the Poynting vector to arbitrary symmetric
energy-momentum tensors.

For Maxwell electrodynamics, this identification reproduces the familiar
expressions
\begin{equation}
\mathbf{J}_{0}(x)
=
-\mathbf{x}\times(\mathbf{E}\times\mathbf{B}),
\qquad
\mathbf{S}_{0}(x)
=
\mathbf{E}\times\mathbf{B},
\label{eq:S0-def}
\end{equation}
in agreement with the Belinfante construction.

In the presence of RGUP-induced higher-derivative operators, the symmetric
Belinfante energy-momentum tensor acquires additional contributions beyond the
Maxwell term. As shown in the previous sections, the resulting angular-momentum
density admits a natural decomposition,
\begin{equation}
\mathbf{J}_{\mathrm{Bel}}(x)
=
\mathbf{J}_{0}(x)
+
\mathbf{J}_{1}(x)
+
\mathbf{J}_{2}(x)
+
\mathbf{J}_{3}(x),
\label{eq:JBel-decomp}
\end{equation}
where
\begin{align}
\mathbf{J}_{0}(x)
&=
\mathbf{x}\times(\mathbf{E}\times\mathbf{B}),
\label{eq:J0}\\[4pt]
\mathbf{J}_{1}(x)
&=
\gamma\, \mathbf{x}\times\!\left[(\Box\mathbf{E})\times\mathbf{B}\right],
\label{eq:J1}\\[4pt]
\mathbf{J}_{2}(x)
&=
\gamma\, \mathbf{x}\times\!\left[\mathbf{E}\times(\Box\mathbf{B})\right],
\label{eq:J2}\\[4pt]
\mathbf{J}_{3}(x)
&=
2\gamma\,\mathbf{x}\times
\Big[
E^{a}\,\partial^{0}\partial_{0}\nabla A_{a}
-
E^{a}\,\partial^{0}\partial_{a}\nabla A_{c}
\nonumber \\
& \qquad \qquad \qquad \qquad+
\epsilon^{abm}B^{m}\,\partial^{0}\partial_{a}\nabla A_{b}
\Big].
\label{eq:J3}
\end{align}

%\begin{align}\mathbf{J}_{0}(x)&=-\,\mathbf{x}\times(\mathbf{E}\times\mathbf{B}),\label{eq:J0}\\[3pt]\mathbf{J}_{1}(x)&=-\,\gamma\, \mathbf{x}\times\!\left[(\Box\mathbf{E})\times\mathbf{B}\right],\label{eq:J1}\\[3pt]\mathbf{J}_{2}(x)&=-\,\gamma\, \mathbf{x}\times\!\left[\mathbf{E}\times(\Box\mathbf{B})\right],\label{eq:J2}\\[3pt]\mathbf{J}_{3}(x)&=2\gamma\, \mathbf{x}\times\!\left[E^{a}\,\partial^{0}\partial_{0}\nabla A_{a}-E^{a}\,\partial^{0}\partial_{a}\nabla A_{0}\\[3pt]\qquad \qquad \qquad \qquad+\epsilon^{abm}B^{m}\,\partial^{0}\partial_{a}\nabla A_{b}\right].\label{eq:J3}\end{align}

Each contribution is manifestly of the orbital form
$\mathbf{J}_{n}(x)=-\,\mathbf{x}\times\mathbf{S}_{n}(x)$, which allows one to define a generalized RGUP-modified Poynting vector,
\begin{equation}
\mathbf{S}_{\mathrm{RGUP}}(x)
=
\mathbf{S}_{0}(x)
+
\mathbf{S}_{1}(x)
+
\mathbf{S}_{2}(x)
+
\mathbf{S}_{3}(x),
\label{eq:SRGUP-def}
\end{equation}
with
\begin{align}
\mathbf{S}_{0}(x)
&=
\mathbf{E}\times\mathbf{B},
\label{eq:S0}\\[4pt]
\mathbf{S}_{1}(x)
&=
\gamma\,(\Box\mathbf{E})\times\mathbf{B},
\label{eq:S1}\\[4pt]
\mathbf{S}_{2}(x)
&=
\gamma\,\mathbf{E}\times(\Box\mathbf{B}),
\label{eq:S2}\\[4pt]
\mathbf{S}_{3}(x)
&=
-\,2\gamma\,
\!\Big[
E^{a}\,\partial^{0}\partial_{0}\nabla A_{a}
-
E^{a}\,\partial^{0}\partial_{a}\nabla A_{c}
\nonumber\\ 
&\qquad \qquad \qquad +
\epsilon^{abm}B^{m}\,\partial^{0}\partial_{a}\nabla A_{b}
\Big].
\label{eq:S3}
\end{align}

The Belinfante relation then takes the form
\begin{equation}
\mathbf{J}_{\mathrm{Bel}}(x)
=
-\,\mathbf{x}\times\mathbf{S}_{\mathrm{RGUP}}(x),
\label{eq:J-SRGUP}
\end{equation}
and, owing to the symmetry of the Belinfante energy-momentum tensor,
\begin{equation}
S^{i}_{\mathrm{RGUP}}(x)
=
\mathscr{T}^{0i}(x),
\label{eq:S=T0i}
\end{equation}
so, $\mathbf{S}_{\mathrm{RGUP}}$ indeed represents the momentum density of the
RGUP-corrected Gauge field. The modified
Poynting vector explicitly shows how the minimal-length effects redistribute the
flow of energy and momentum in the gauge sector, while preserving gauge
invariance and the structure of the angular momentum density.

\section{Conclusion}\label{sec6}
The existence of a minimal measurable length is a common consequence of several approaches to quantum gravity. Such minimal length effects are naturally encoded through GUP and its Lorentz-invariant extension known as RGUP, via deformed commutation relations.These modifications alter the underlying phase-space structure, effectively acting as a soft ultraviolet regulator and providing a systematic low-energy description of quantum-gravitational corrections well below the Planck scale. It is therefore of fundamental interest to understand how these deformations manifest themselves in quantum field theory, particularly in gauge dynamics.

In this work, we have developed a systematic and consistent formulation of the canonical and Belinfante EMT, together with the associated angular momentum structures, for gauge fields in the presence of RGUP. The modified phase-space structure induces higher-derivative corrections in the gauge field Lagrangian, from which we explicitly derived the RGUP-corrected canonical and symmetric (Belinfante) EMTs. Using these results, we constructed the corresponding angular momentum currents and identified both the orbital and intrinsic spin contributions of the gauge field under Planck-scale deformations.

We showed that, as in standard gauge field theory, the canonical EMT in the RGUP framework is neither symmetric nor gauge invariant. Thus, using Noether's second theorem as shown in \cite{freese2022noether}, we obtained symmetric and gauge-invariant EMT, allowing for a Lorentz-covariant definition of the total angular momentum current. The resulting expressions for the orbital and spin angular momentum reduce smoothly to their conventional quantum electrodynamics counterparts in the limit of vanishing RGUP deformation, ensuring consistency with the standard theory.

A central result of our analysis is that the fundamental conservation laws of relativistic field theory remain intact within the RGUP framework. Although the orbital and spin components of the canonical angular momentum are not conserved separately, the total angular momentum current constructed from the Belinfante tensor is conserved, in accordance with Lorentz symmetry and Noether’s theorem. This demonstrates that RGUP provides a consistent kinematical deformation of gauge dynamics that preserves the underlying symmetry structure of relativistic gauge field theory.
We further extended our analysis to derive the RGUP-modified momentum density, or Poynting vector, describing the flow of electromagnetic energy in the presence of minimal-length corrections. This result establishes a direct connection between microscopic RGUP-induced modifications of gauge-field kinematics and macroscopic energy and momentum transport.

In summary, this work presents a comprehensive analysis of the angular momentum structure of gauge fields in the RGUP framework, deriving modified canonical and Belinfante energy-momentum tensors, angular momentum currents, and the Poynting vector while preserving fundamental conservation laws and Lorentz covariance. The RGUP corrections introduce higher-derivative contributions that enrich local densities and momentum transport, offering a concrete link between Planck-scale kinematics and macroscopic electromagnetic phenomena, with the standard Maxwell limit recovered as $\gamma \rightarrow 0$.
While the present framework is robust, internally consistent, and provides the first fully covariant treatment of gauge-field angular momentum under Lorentz-invariant minimal-length deformations, we acknowledge that it currently lacks quantitative predictions of observable deviations, such as explicit upper bounds on the RGUP parameter $\gamma$ derived from high-precision electromagnetic data (atomic transition rates, CMB polarization anisotropies and  ultra-precise polarization measurements in astrophysical sources) \cite{Eskilt:2022cff,shao2011lorentz,kostelecky2009electrodynamics}. Likewise, the analysis is restricted to free gauge fields in flat Minkowski spacetime and does not yet include coupling to charged matter fields or gravitational backreaction, both of which are expected to introduce additional structure and potentially amplify or suppress the RGUP-induced modifications in realistic physical settings. We are strongly committed to closing these gaps in forthcoming extensions of this work. We plan to derive concrete phenomenological bounds on $\gamma$ using existing and upcoming experimental constraints, perform detailed numerical simulations of the modified angular-momentum currents in interacting QED-like systems, and explore astrophysical and laboratory signatures that can be unambiguously distinguished from standard quantum electrodynamics. These steps will transform the present theoretical foundation into a fully testable minimal-length phenomenology and significantly strengthen the impact of this line of research.

\appendix
\section{Total Variations of Gauge Fields Under local Transformation}\label{appa}

We consider a local transformation
\begin{equation}
x'^{\mu} = x^{\mu} + \xi^{\mu}(x),
\end{equation}
where $\xi^\mu(x)$ is infinitesimal. Only leading order terms in $\xi^\mu$ and its first order derivatives are retained. The total derivative of $A_\mu$ is defined as
\begin{equation}
\Delta A_\mu(x) = A'_\mu(x') - A_\mu(x),
\end{equation} 
Since $A_\mu$ is a covariant vector field,
\begin{equation}
A'_{\mu}(x')
=
\frac{\partial x^\alpha}{\partial x'^\mu}
A_\alpha(x),\quad \text{using definition,}\quad \frac{\partial x^\alpha}{\partial x'^\mu}
=
\delta^\alpha_{\mu}
-
\partial_{\mu}\xi^\alpha
\end{equation}
gives 
\begin{equation}
\Delta A_\mu
=
-
(\partial_\mu \xi^\alpha)\,A_\alpha.
\end{equation}
Similarly, the total variation for $\partial_\nu A_\mu$ is 
\begin{equation}
\Delta(\partial_\nu A_\mu)
=
\partial'_\nu A'_\mu(x')
-
\partial_\nu A_\mu(x).
\end{equation}
After, considering the first-order derivative and linear $\xi$, we obtain
\vspace{-0.9mm}
\begin{equation}
\Delta(\partial_\nu A_\mu)
=
-
(\partial_\nu \xi^\alpha)\,\partial_\alpha A_\mu
-
(\partial_\mu \xi^\alpha)\,\partial_\nu A_\alpha.
\end{equation}
For third order derivative term $\partial_\rho \partial^\rho \partial_\nu A_\mu$, we define
\begin{equation}
\Delta(\partial_\rho \partial^\rho \partial_\nu A_\mu)
=
\partial'_\rho \partial'^\rho \partial'_\nu A'_\mu
-
\partial_\rho \partial^\rho \partial_\nu A_\mu,
\end{equation}
now, converting the primes to unprime coordinates
\begin{align}
\Delta\!\left(\partial_\rho \partial^\rho \partial_\nu A_\mu\right)
=&\;
\Bigl(\partial_\rho \partial^\rho
- \partial_\sigma \xi^\rho \partial_\rho \partial^\sigma
- \partial_\rho \xi^\alpha \partial_\alpha \partial^\rho \Bigr)
\nonumber\\
&
\Bigl(\partial_\nu
- \partial_\nu \xi^\beta \partial_\beta \Bigr)
\Bigl(A_\mu
- \partial_\mu \xi^{a} A_{a} \Bigr)
- \partial_\rho \partial^\rho \partial_\nu A_\mu .
\end{align}

\begin{align}
\Delta\!\left(\partial_\rho \partial^\rho \partial_\nu A_\mu\right)
=&\;
\Bigl(\partial_\rho \partial^\rho
- \partial_\sigma \xi^\rho \partial_\rho \partial^\sigma
- \partial_\rho \xi^\alpha \partial_\alpha \partial^\rho \Bigr)
\nonumber\\
&
\Bigl(\partial_\nu
- \partial_\nu \xi^\beta \partial_\beta \Bigr)
\Bigl(A_\mu
- \partial_\mu \xi^{a} A_{a} \Bigr)
- \partial_\rho \partial^\rho \partial_\nu A_\mu .
\end{align}
we get,
\begin{align}
\Delta(\partial_\rho \partial^\rho \partial_\nu A_\mu)
=&
-
(\partial_\mu \xi^a)\,
\partial^\rho \partial_\rho \partial_\nu A_a
-
(\partial_\nu \xi^\beta)\,
\partial^\rho \partial_\rho \partial_\beta A_\mu
\nonumber\\
&
-
(\partial_\sigma\xi^\rho)\,
\partial_\rho \partial^\sigma \partial_\nu A_\mu
-
(\partial_\rho\xi^\alpha)\,
\partial_\alpha \partial^{\rho} \partial_\nu A_\mu.
\end{align}

\bibliographystyle{unsrt}
\bibliography{ref}

%\begin{thebibliography}{99}

%\end{thebibliography}

\end{document}